\documentclass[12pt]{article}
\usepackage{amssymb,epsfig}

\renewcommand{\thefootnote}{\fnsymbol{footnote}}

\begin{document}

\title{
\begin{flushright}
\ \\*[-80pt] 
\begin{minipage}{0.2\linewidth}
\normalsize
arXiv:0707.2671 \\
YITP-07-43 \\
TU-794 \\
KUNS-2087 \\*[50pt]
\end{minipage}
\end{flushright}
{\Large \bf 
More about F-term uplifting
\\*[20pt]}}

\author{Hiroyuki~Abe$^{1,}$\footnote{
E-mail address: abe@yukawa.kyoto-u.ac.jp}, \ 
Tetsutaro~Higaki$^{2,}$\footnote{
E-mail address: tetsu@tuhep.phys.tohoku.ac.jp}, \ and \ 
Tatsuo~Kobayashi$^{3,}$\footnote{
E-mail address: kobayash@gauge.scphys.kyoto-u.ac.jp}\\*[20pt]
$^1${\it \normalsize 
Yukawa Institute for Theoretical Physics, Kyoto University, 
Kyoto 606-8502, Japan} \\
$^2${\it \normalsize 
Department of Physics, Tohoku University, 
Sendai 980-8578, Japan} \\
$^3${\it \normalsize 
Department of Physics, Kyoto University, 
Kyoto 606-8502, Japan} \\*[50pt]}

\date{
\centerline{\small \bf Abstract}
\begin{minipage}{0.9\linewidth}
\medskip 
\medskip 
\small
We study moduli stabilization and a realization of de Sitter 
vacua in generalized F-term uplifting scenarios of the KKLT-type 
anti-de Sitter vacuum, where the uplifting sector $X$ directly 
couples to the light K\"ahler modulus $T$ in the superpotential 
through, e.g., stringy instanton effects. F-term uplifting can 
be achieved by a spontaneous supersymmetry breaking sector, e.g., 
the Polonyi model, the O'Raifeartaigh model and the 
Intriligator-Seiberg-Shih model. Several models with the $X$-$T$ 
mixing are examined and qualitative features in most models {\it 
even with such mixing} are almost the same as those in the KKLT scenario. 
One of the quantitative changes, which are relevant to the phenomenology, 
is a larger hierarchy between the modulus mass $m_T$ and the gravitino 
mass $m_{3/2}$, i.e., $m_T/m_{3/2} = {\cal O}(a^2)$, where $a \sim 4 \pi^2$. 
In spite of such a large mass, the modulus F-term is suppressed not 
like $F^T = {\cal O}(m_{3/2}/a^2)$, but like $F^T = {\cal O}(m_{3/2}/a)$ 
for $\ln (M_{Pl}/m_{3/2}) \sim a$, because of an enhancement factor 
coming from the $X$-$T$ mixing. Then we typically find a mirage-mediation 
pattern of gaugino masses of ${\cal O}(m_{3/2}/a)$, while the scalar 
masses would be generically of ${\cal O}(m_{3/2})$. 
\end{minipage}
}

\begin{titlepage}
\maketitle
\thispagestyle{empty}
\end{titlepage}


\renewcommand{\thefootnote}{\arabic{footnote}}
\setcounter{footnote}{0}

\section{Introduction}
It is quite important to realize the real world 
based on string theory~\cite{Green:1987sp}, though we have lots 
of difficulties to be overcome. We do not have a 
definite answer in string theory why the dimension of our spacetime 
is four or a definite scenario to derive the standard model (SM),   
the $SU(3) \times SU(2)\times U(1)$ gauge symmetry, 
three families of quarks and leptons, experimental values of 
gauge couplings and fermion masses and mixing angles, etc. 
Many attempts have been tried to solve these problems in several 
aspects step by step. 

One of the most severe difficulties to be solved, 
is the moduli stabilization problem.
This problem always occurs when one assumes that the observed 
spacetime dimension is realized in string theory. 
Non-perturbative effects, such as gaugino 
condensations~\cite{Dine:1985rz} were used to be
considered for the moduli stabilization~\cite{Nilles:2004zg}.
Recently a new mechanism of moduli stabilization has been brought to 
the attention. That is a compactification with closed string 
flux~\cite{Giddings:2001yu} with orientifold planes and D-branes, 
which can be a source of the fluxes and cause non-perturbative effects. 

Advantages of such flux compactification are that the internal 
compact space can be warped by flux generating large hierarchies, 
and also that we can fix a lot of moduli, e.g. a number of 
${\cal O}(100)$, simultaneously. For example, in type IIB orientifold 
model on Calabi-Yau, the dilaton, complex structure and D7-brane moduli 
are stabilized by an imaginary self dual flux. If our world can be 
described by string theory with D-branes, 
it is natural to consider both fluxes and non-perturbative effects 
as sources of moduli fixing. 

Furthermore, moduli fields can play important roles in the low energy 
phenomenology. Moduli fields determine compactification scales of the 
internal space. Then, the moduli generically couple to four-dimensional 
kinetic terms of gauge fields, and their vacuum expectation values 
(vevs) determine gauge couplings and similarly other
couplings like Yukawa couplings. 
Moreover, F-components of moduli superfields, which are also given by 
nonvanishing vevs of moduli themselves, can be a source of supersymmetry 
(SUSY) breaking and induce soft SUSY breaking terms in the visible
sector. SUSY can be broken 
nonperturbatively through the moduli fixing procedure. Therefore, 
if we find any signatures of SUSY breaking at near future experiments, 
it is very interesting not only from the phenomenological 
viewpoint but also from the viewpoint of string theory.

Recently the authors of~\cite{Kachru:2003aw} proposed a semi realistic 
scenario based on the flux compactification with D-branes and an anti 
D-brane, that is called KKLT scenario. In this scenario, all moduli are 
fixed by flux and non-perturbative effects on D-branes and the 
de Sitter/Minkowski vacuum is realized. 
SUSY can be broken moderately by a red shifted anti D-brane which is 
sitting at the tip of warped throat and is well separated from the light 
modulus as well as the visible sector. Because of this sequestering 
structure, the scalar potential of the modulus, which has a SUSY minimum 
with a negative vacuum energy before adding the uplifting effect, 
can be easily uplifted allowing a tuning 
of the cosmological constant. The minimum becomes a SUSY breaking 
metastable vacuum for closed string sector.

In this scenario, 
F-components of K\"ahler moduli are suppressed compared with the gravitino 
mass~\cite{Choi:2005ge}. This results in the fact that moduli mediation 
and anomaly mediation~\cite{Randall:1998uk} are comparable~\cite{Choi:2005uz}, 
that is, the so-called mirage mediation. 
Thus, the gravitino mass is of ${\cal O}(100)$ TeV 
to realize the low-energy SUSY breaking in this scenario.
This kind of the mediation mechanism causes 
a distinctive pattern of sparticle spectra at the TeV scale
\cite{Choi:2005uz,Endo:2005uy} and naturally 
solves the SUSY CP problem~\cite{Choi:2005ge,Choi:2005uz}, 
though it may have a 
gravitino overproduction problem~\cite{Nakamura:2006uc}. It is also known 
that, with the mirage mediation, the so-called little hierarchy problem 
can be avoided within the minimal SUSY SM (MSSM)~\cite{Choi:2005hd}, 
where the mirage unification of the wino and the gluino masses at 
the TeV scale is important~\cite{Abe:2007kf}.

The source of the uplifting is the anti-brane in the original KKLT 
scenario.
However, it can be replaced by a dynamical SUSY breaking. 
Recently, metastable vacua with dynamical SUSY breaking 
have been studied in field-theoretical model 
building~\cite{Intriligator:2006dd,Forste:2006zc}.
Also, including realization of these field-theoretical models, 
metastable models of a dynamical 
SUSY breaking have been studied 
not only in the closed string sector but also in the open string
sector~\cite{Franco:2006es}. 
By introducing such SUSY breaking 
sector into the KKLT model, we can construct F-term uplifting 
scenarios~\cite{Saltman:2004sn,Dudas:2006gr,Abe:2006xp,Kallosh:2006dv,
Lebedev:2006qc}, where the Polonyi model~\cite{Polonyi:1977pj}, 
the O'Raifeartaigh model~\cite{O'Raifeartaigh:1975pr} and 
the Intriligator-Seiberg-Shih (ISS) model~\cite{Intriligator:2006dd} 
have been considered as the F-term uplifting sector.\footnote{
The ISS model can correspond to the O'Raifeartaigh model after 
integrating out heavy modes in both models.} 
F-term uplifting models are more interesting than other uplifting 
schemes, because the size of SUSY breaking is controllable 
and a small gravitino mass, which is 
comparable to the electroweak scale, can be realized. 
In the D-term uplifting~\cite{Burgess:2003ic}\footnote{
See also~\cite{Choi:2006bh}.} and 
the K\"ahler uplifting~\cite{Westphal:2006tn} schemes, 
to control the size of SUSY breaking is not simple, 
and they would  naturally lead to a large SUSY breaking scale, 
which is comparable to the Planck scale. 

In the F-term uplifting scenario, we find slight differences from 
the original KKLT predictions with the anti-brane, although the 
qualitative features are not changed. For example, the ratio of the 
anomaly mediation to the modulus mediation takes a different value 
depending on the model and as a consequence the prediction of 
sparticle mass spectra at a low energy scale is different from the 
original KKLT scenario. 

In this paper, we generalize the F-term uplifting scenarios such that 
the uplifting sector $X$ directly couples to the light K\"ahler modulus 
$T$ in the nonperturbative superpotential induced by, e.g., stringy 
instanton effects. We will show that in most cases the qualitative 
features of the KKLT scenario can still be almost the same even with the 
modulus mixing to the uplifting sector. One of the quantitative changes, 
which would be phenomenologically relevant, is a larger hierarchy between 
the modulus mass and the gravitino mass than one in the original 
KKLT model. We will typically find mirage-mediation type gaugino masses, 
while the scalar masses are of the order of the gravitino mass in general. 

We arrange the sections of this paper as follows. 
In Sec.~\ref{sec:flift}, we study the Polonyi-KKLT model and 
the ISS-KKLT model as concrete 
examples of the F-term uplifting. We introduce a mixing 
between the light modulus $T$ and the uplifting sector $X$ in the 
superpotential, and find the minimum of the scalar potential based 
on the perturbation from a reference point where 
both the Polonyi/ISS-type structure and the KKLT-type structure 
would be realized for $X$ and $T$, respectively. 
Here we assume that constants of the Polonyi/ISS models are (dynamically) 
generated by flux or stringy instantons~\cite{Florea:2006si} 
depending on the light modulus\footnote{For Affleck-Dine-Seiberg-KKLT 
model~\cite{Affleck:1983mk}, see \cite{Acharya:2007rc}.}. 
Then, in Sec.~\ref{sec:orderparameter}, we analyze the SUSY breaking 
order parameters and the masses of light modes. After deriving some 
general expressions in Sec.~\ref{ssec:gen}, 
we show some results in several concrete examples 
in Sec.~\ref{ssec:ex:pkklt} and \ref{ssec:ex:issrt}. 
Sec.~\ref{sec:conclusion} is devoted to conclusions and discussions.

\section{Moduli involved F-term uplifting}
\label{sec:flift}
For concreteness, we consider type IIB supergravity with 
the dilaton $S$, complex moduli $U$ and a single K\"ahler modulus $T$. 
Our analysis can be extended to the case with several K\"ahler moduli.
Here and hereafter we use the mass unit with $M_{Pl}=1$, 
where $M_{Pl}$ is the 
four-dimensional Planck mass.
As in the original KKLT model~\cite{Kachru:2003aw}, we assume that the 
dilaton $S$ and complex structure moduli $U$ 
are stabilized by the flux induced superpotential 
\begin{eqnarray}
W_{\rm flux}(S,U)=\int G_3 \wedge \Omega, 
\nonumber 
\end{eqnarray}
because of large supersymmetric masses 
$\langle \partial_S \partial_U W_{{\rm flux}}\rangle \sim 1$~\cite{Abe:2006xi}.
Then, the dilaton and the 
complex structure moduli are much heavier than the K\"ahler modulus $T$, 
and these do not affect the low energy dynamics of $T$ and the 
visible sector.

In the KKLT model, the K\"ahler modulus $T$ is stabilized 
by a nonperturbative effect on a $D7$-brane, that is, the 
following superpotential is considered,
\begin{eqnarray}
W_{\rm KKLT}=w_0-Ae^{-aT}, 
\nonumber
\end{eqnarray} 
with the K\"ahler potential $-3 \ln (T+\bar{T})$, 
where the constant $w_0$ in the superpotential originates from the flux 
induced superpotential $W_{\rm flux}$ or a gaugino condensation which 
depend on heavy moduli such as $S$~\cite{Abe:2005rx}, and 
the second term is due to the nonperturbative effect. 
The potential minimum corresponds to a SUSY AdS vacuum.
In order to uplift the AdS minimum to a Minkowski one, 
anti $D3$-branes are introduced at the tip of the warped throat, 
which is well sequestered from the K\"ahler modulus $T$ as well as 
the visible sector, and the effect of anti $D3$-branes just 
appears as an ($T$-dependent) uplifting potential, which is an 
explicit SUSY breaking term in terms of the $N=1$ supergravity, 
added to the standard F-term scalar potential of $T$.

Instead of adding such anti D-branes, 
we can uplift the SUSY AdS minimum by adding a superpotential 
term $W_{\rm lift}(T,X)$, which leads to a nonvanishing 
F-term of the hidden sector field $X$.
That is the F-term uplifting. 
In the original F-term uplifting scenarios~\cite{Saltman:2004sn,
Dudas:2006gr,Abe:2006xp,Kallosh:2006dv,Lebedev:2006qc}, 
the uplifting sector is basically assumed to be well separated from the 
light modulus
 as anti $D3$-branes in the KKLT scenario, 
i.e. $\partial W_{\rm lift}(T,X)/\partial T = 0$. 
As the spontaneous SUSY breaking sector $X$, 
the Polonyi model, O'Raifeartaigh model and ISS model have 
been considered.
Here, we study generalized F-term uplifting scenarios where 
the hidden sector field $X$, which is responsible for the nonvanishing 
F-term, is directly couples to the light K\"ahler modulus due to, e.g., 
stringy instanton effects~\cite{Florea:2006si}. 
Such instanton effects for instance induce a mass term or a tadpole term of 
the hidden sector field $X$ which depends on the light K\"ahler modulus. 
Similar situation has been studied based on heterotic and M-theoretical
models with 
multiple light moduli~\cite{Acharya:2007rc,Serone:2007sv}\footnote{
See also~\cite{Abe:2005pi} for a model with similar properties.}. 

We will show that the F-term uplifting is still valid in most cases without 
changing the qualitative features of both the light modulus (KKLT) 
sector and the uplifting (dynamical SUSY breaking) sector.

\subsection{Polonyi-KKLT model}
One of the simplest models for the F-term uplifting is the Polonyi-KKLT 
model~\cite{Abe:2006xp,Lebedev:2006qc,Dine:2006ii}. 
The K\"ahler potential $K$ and the superpotential $W$ are given by 
\begin{eqnarray}
K &=& \Omega(T,\bar{T})+Z(T,\bar{T})|X|^2, 
\label{eq:kahler:pkklt} \\
W &=& w_0-Ae^{-aT}+Be^{-bT}X, 
\label{eq:super:pkklt}
\end{eqnarray}
where $\Omega(T,\bar{T})$ is the K\"ahler potential of overall volume 
(K\"ahler) modulus, which is typically given by $-3 \ln (T+\bar{T})$, 
and $Z(T,\bar{T})=K_{X\bar{X}}$ is the K\"ahler metric of $X$. 
In the superpotential, we assume a typical magnitude of parameters 
$$|a| ,\ |b| \ \sim \ 4 \pi^2.$$
When $A=0$ and $b=0$, the above superpotential corresponds 
to the Polonyi model, i.e. $W_{\rm Polonyi}=w_0+BX$,  
and leads to spontaneous SUSY breaking with 
nonvanishing $F^X$.
The third term in the right hand side is the mixing between 
$X$ and $T$, and such a mixing can be induced by 
sting instanton effects.
The Polonyi-KKLT model without $X$-$T$ mixing, i.e. $b=0$, has 
been studied in~\cite{Abe:2006xp,Lebedev:2006qc,Dine:2006ii}.  

{}From these K\"ahler potential and superpotential, 
we can derive the $F$-term scalar potential $V$ 
using the standard $N=1$ supergravity formula: 
\begin{eqnarray}
V &=& e^G (G^{I\bar{J}}G_IG_{\bar{J}}-3) 
\ = \ K_{I\bar{J}}F^I \bar{F}^{\bar{J}}-3e^K|W|^2, 
\nonumber \\
G &=& K + \ln|W|^2, \qquad 
F^I \ = \ -e^{K/2}K^{I\bar{J}}D_{\bar{J}} \bar{W}, \qquad 
D_IW \ = \ W_I+K_IW. 
\label{eq:gkahler:pkklt}  
\end{eqnarray}

We try to find a minimum of the potential by a perturbation from 
the reference point. We choose the reference point $(X,T)=(X_0,T_0)$ 
which satisfies the following conditions\footnote{
Here, because the number of complex parameters in superpotential 
is less than four, we can always make the vevs of fields real by 
field redefinitions (shifts or rotations).}, 
\begin{eqnarray}
V_X \big|_0 &=& V \big|_0 \ = \ 0, \qquad 
D_TW \big|_0 \ = \ 0,\qquad 
X_0 \ = \ \bar{X}_0, \qquad 
T_0 \ = \ \bar{T}_0. \qquad 
\label{eq:refpt}
\end{eqnarray}
At this reference point, the KKLT-like modulus property 
and the Polonyi-like SUSY breaking property would be realized.
We tune our parameters to obtain 
almost vanishing vacuum energy $V=0$. In this Polonyi-KKLT model, 
the reference point (\ref{eq:refpt}) is characterized by 
\begin{eqnarray}
X_0 &\sim& T_0 
\ \sim \ \partial_T^n \partial_{\bar{T}}^m  \Omega \big|_0 
\ \sim \ \partial_T^n \partial_{\bar{T}}^m  Z \big|_0 
\ \sim \ {\cal O}(1), 
\nonumber \\
W \big|_0 &\sim& W_T \big|_0=-K_TW \big|_0, \qquad 
|\partial_T^{n+2} W| \big|_0 \ \sim \ a^{n+2}|W|, 
\label{eq:refpt:ftr}
\end{eqnarray}
where $n,m=0,1,2,\ldots$. 

The true minimum of the potential is denoted by 
$$\langle \Phi^I \rangle \ = \ \Phi^I \big|_0 + \delta \Phi^I,$$
where $\Phi^I=(X,\,T)$. 
Assuming $\delta \Phi^I/\Phi^I \big|_0 \ll 1$, 
we expand $V_I$ as 
\begin{eqnarray}
V_I &=& V_I \big|_0 
+ V_{I\bar{J}} \big|_0 \delta \Phi^{\bar{J}} 
+ V_{IJ} \big|_0 \delta \Phi^J 
+{\cal O}(\delta \Phi^2)
\nonumber \\
&=& V_I \big|_0 
+ \hat{V}_{IJ} \big|_0 \delta \Phi^J 
+{\cal O}(\delta \Phi^2), 
\nonumber
\end{eqnarray}
where $\hat{V}_{IJ}=V_{I\bar{J}} + V_{IJ}$ 
and $\delta \Phi^{\bar{I}} = \delta \Phi^I$. 
The stationary condition $V_I=0$ results in 
$$\delta \Phi^I \ = \ -\hat{V}^{IJ} V_J \big|_0 
+{\cal O}(\delta \Phi^2),$$ 
where $\hat{V}_{IJ} \hat{V}^{JK} \big|_0 = \delta_I^{\ K}$. 
For $ V_{I\bar{J}} > V_{IJ}$, we find~\cite{Lebedev:2006qc}
\begin{eqnarray}
\delta T &\simeq& \left. 
\frac{V_{\bar{T}}}{
|V_{X\bar{T}}|^2/V_{X\bar{X}}
-V_{T\bar{T}}} \right|_0, \qquad 
\delta X \ \simeq \ \left. 
-\frac{V_{T\bar{X}}}{V_{X\bar{X}}} \right|_0 \delta T. 
\nonumber
\end{eqnarray}

First we consider the case that the K\"ahler mixing is small, 
$\big| Z^{-1}\partial_TZ \big| < {\cal O}(a^{-1})$ at the 
reference point. By estimating the orders of 
$V_I \big|_0$, $V_{I\bar{J}} \big|_0$ and $V_{IJ} \big|_0$, 
we find $X_0 = \sqrt{3}-1$, $T_0 ={\cal O}(1)$, 
$$V_T \big|_0 \ \sim \ b \, e^G|_0,$$
and 
\begin{eqnarray}
\hat{V}_{IJ} \big|_0 
&\sim& \left( 
\begin{array}{cc}
V_{X\bar{X}} & V_{X\bar{T}} \\
V_{T\bar{X}} & V_{T\bar{T}} 
\end{array} \right)
\nonumber
\sim \left( 
\begin{array}{cc}
b^2 +1 & b(m_T/m_{3/2}+1) \\
b(m_T/m_{3/2} + 1) & m_{T}^2/m_{3/2}^2 + b^2 
\end{array} \right)m_{3/2}^2 \\
&\sim& \left( 
\begin{array}{cc}
b^2 + 1 & ab^2+b \\
ab^2+b & a^2b^2 + b^2 
\end{array} \right)m_{3/2}^2, 
\nonumber
\end{eqnarray}
where $m_{3/2}=e^{G/2}|_0$ and 
\begin{eqnarray}
m_T &=& 
-e^{K/2}K^{T\bar{T}}W_{TT} \Big|_0 
\ \sim \ ab(T_0+\bar{T_0})^2m_{3/2}. 
\label{eq:mtpkklt}
\end{eqnarray}

Then, for $a \sim b \ne 0$, we find that the Hessian 
can be positive, and obtain 
\begin{eqnarray}
\frac{\delta T}{T_0} &\simeq& \frac{1}{b^3} \ \ll \ 1, \qquad 
\frac{\delta X}{X_0} \ \simeq \ \frac{1}{b^2} \ \ll \ 1.
\nonumber
\end{eqnarray}
This result should be compared with 
$\delta T/T_0 \simeq a^{-2}$ and 
$\delta X/X_0 \simeq a^{-1}$ derived in the case that $Be^{-bT}$ 
is replaced by a $T$-independent constant in the superpotential. 
(When one sets $b=1$ 
in the above expressions, the results of such case are obtained.) 
This difference originates from the fact that the superpotential of 
modulus is not effectively a KKLT type but rather 
a {\it racetrack} type because of 
$ \langle X \rangle = {\mathcal O}(1)$. From this result, we find that 
the true minimum resides in a perturbative region from the reference 
point (\ref{eq:refpt}). Then, we expect that the SUSY breaking 
structure of the Polonyi-KKLT model is not affected qualitatively by 
the mixing between the Polonyi sector $X$ and the KKLT sector $T$, 
although quantitatively the modulus mass becomes the racetrack-type 
(\ref{eq:mtpkklt}) for $b \sim a$~\cite{Choi:2005ge}. 

Finally we comment that a considerable K\"ahler mixing 
$\big| Z^{-1} \partial_T Z \big| = {\cal O}(1)$ 
at the reference point would affect the above order estimations. 
For example, we obtain 
$V_T|_0 \sim  b (1+ aX_0 \partial_T Z )e^{G}|_0  
\sim b^2e^G|_0$ and $\delta T/T_0 \sim 1/b^2,~ \delta X/X_0 \sim 1/b$, 
assuming that $T_0$ and $X_0$ are of ${\cal O}(1)$.
Then without a tuning between $a$ and $b$, we may find 
\begin{eqnarray}
\nonumber
F^X &\sim& D_XW = D_XW|_0 + W_{XT}|_0 \delta T + K_XW_X|_0 \delta X 
+ \cdots \\
\nonumber
&\sim & m_{3/2}\left( 1 + \frac{1}{b} + \cdots \right) ,\\
\nonumber
F^T &\sim& D_{T}W = W_{TT}|_0\delta T + W_{TX}|_0 \delta X + W_{XTT}|_0\delta X \delta T + 
\frac{1}{2}W_{TTT}|_0(\delta T)^2 
+ \cdots \\
\nonumber
&\sim & m_{3/2} \left( 1 + \frac{1}{b} +\cdots \right), 
\end{eqnarray}
where we wrote both $1/a$ and $1/(a-b)$ as $1/b$. 
If $a-b$ is of ${\cal O}(1)$, the above expansion in some cases may not 
converge, and the perturbation may be invalid as in the ISS-KKLT model 
shown later. For the case in which it converges ($a-b={\cal O}(4\pi^2)$ etc.), 
we leave a concrete study as a future work.

\subsection{ISS-KKLT model}
Another interesting source of uplifting is 
the ISS model~\cite{Intriligator:2006dd} 
and in this subsection we consider the 
ISS-KKLT model~\cite{Dudas:2006gr,Abe:2006xp}. 
After heavy modes are integrated out around the SUSY 
breaking minimum, the ISS model leads to the same superpotential 
as the Polonyi model, where the field $X$ corresponds to 
the meson field and the tadpole term of $X$ corresponds to 
a mass term of quarks $q$ and $\bar q$ in the dual side, 
i.e. $X \sim q \bar q$.
A $T$-dependent mass term of $q$ and $\bar q$  can be 
generated by string instanton effect like $q \bar q e^{-bT}$.
That can be an origin of the third term of the right hand side 
in Eq.~(\ref{eq:super:pkklt}).
At any rate, we use the same superpotential as Eq.~(\ref{eq:super:pkklt}).
However, the K\"ahler 
potential of $X$ receives a one-loop correction from the heavy modes. 
The relevant part can be written as~\cite{Kallosh:2006dv,Kitano:2006wz}
\begin{eqnarray}
{\cal K} &=& K
-\frac{1}{\Lambda^2}Z^{(1)}(T,\bar{T})|X|^4, 
\nonumber
\end{eqnarray}
where $K$ is the tree level K\"ahler potential 
given by Eq.~(\ref{eq:kahler:pkklt}). 
We assume the same superpotential (\ref{eq:super:pkklt}) 
as before\footnote{
If we really adopt the ISS model itself, $\Lambda$ would be dependent 
of $T$ like $\Lambda ^2 \sim 16\pi^{2}Be^{-bRe(T)}$ because the mass 
of heavy modes is also given by $Be^{-bT}$. However we consider $\Lambda$ 
as a constant which comes from the vevs of heavy moduli, that is, 
the ``ISS model'' represents some proper O'Raifeartaigh model in this paper. 
Note that, even if $\Lambda$ is dependent of $T$, the following results 
would not be changed as far as the ISS-like vacuum is stable.}. 
The O'Raifeartaigh model leads to the same K\"ahler potential and 
superpotential after heavy modes are integrated out. 

The scalar potential is then written as 
\begin{eqnarray}
V &=& e^{\cal G} ({\cal G}^{I\bar{J}}{\cal G}_I{\cal G}_{\bar{J}}-3), 
\nonumber \\
&=& e^G (G^{I\bar{J}}G_IG_{\bar{J}}-3) 
+m_X^2(T,\bar{T}) |X|^2 +\cdots, 
\nonumber
\end{eqnarray}
where ${\cal G}={\cal K} + \ln|W|^2$, 
$G$ is given by Eq.~(\ref{eq:kahler:pkklt}) and 
\begin{eqnarray}
m_X^2(T,\bar{T}) 
&=& \frac{4B^2}{\Lambda^2}
e^{-b(T+\bar{T})}\frac{Z^{(1)}(T,\bar{T})}{Z(T,\bar{T})}. 
\nonumber
\end{eqnarray}

We execute a similar analysis to the Polonyi-KKLT model based on the 
reference point (\ref{eq:refpt}) with the order estimation 
(\ref{eq:refpt:ftr}) but now 
$X_0 \simeq (\sqrt{3Z}Z/6Z^{(1)}) \Lambda^2 \sim 10^2 m_{3/2}\ll 1$. 
The important difference between the Polonyi uplifting scenario 
and the ISS uplifting scenario is the size of $X_0$, that is, 
we have $X_0={\cal O}(1)$ in the Polonyi uplifting scenario, while 
$X_0$ is much smaller in the ISS uplifting scenario.
Because of the smallness of $X_0$, the following results are not affected by 
the K\"ahler mixing $\partial_T Z \ne 0$.
In this case, we can estimate the order of $V_I \big|_0$, 
$V_{I\bar{J}} \big|_0$ and $V_{IJ} \big|_0$, and find 
\begin{eqnarray}
V_T \big|_0 &\sim& b\,e^G|_0, 
\nonumber
\end{eqnarray}
and 
\begin{eqnarray}
\nonumber
\hat{V}_{IJ} \big|_0 
&\sim& \left( 
\begin{array}{cc}
V_{X\bar{X}} & V_{X\bar{T}} \\
V_{T\bar{X}} & V_{T\bar{T}} 
\end{array} \right)\,
\sim \left( 
\begin{array}{cc}
1/X_0+b^2 & b(m_T/m_{3/2}+1) \\
b(m_T/m_{3/2} + 1) & m_{T}^2/m_{3/2}^2 + b^2 
\end{array} \right)m_{3/2}^2\\
&\sim& \left( 
\begin{array}{cc}
1/X_0 & ab \\
ab &  a^2+b^2
\end{array} \right)m_{3/2}^2 \,, 
\nonumber
\end{eqnarray}
where $m_{3/2}=e^{G/2}|_0$ and 
$m_T=-e^{K/2}K^{T\bar{T}}W_{TT} \Big|_0 \sim a(T_0+\bar{T_0})m_{3/2}$. 
Because $X_0$ is much smaller than the Planck scale, the effective 
superpotential for the modulus is not the racetrack type but the 
KKLT type. Then, for $a \sim b \ne 0$, we obtain 
\begin{eqnarray}
\delta T/T_0 &\simeq& a^{-1} \ \ll \ 1, \qquad 
\delta X/X_0 \ \simeq \ a \ \gg \ 1.
\nonumber
\end{eqnarray}
This results should be compared with the results  
$\delta T/T_0 \simeq a^{-2}$ and 
$\delta X/X_0 \simeq a^{-1}$ for the case that 
$Be^{-bT}$ is replaced by a $T$-independent constant. 
(When one sets $b=1$, one can obtain the result for such case.) 
We conclude that, with the superpotential mixing between the ISS 
sector and the KKLT sector, the reference point (\ref{eq:refpt}) is far 
from the true minimum.  Then, the SUSY breaking structure would be 
quite different from the ISS-KKLT model without the $X$-$T$ mixing.

\subsection{ISS-racetrack model}
We can evade the problem in the previous subsection 
by adding another nonperturbative effect, 
$Ce^{-cT}(c \neq a)$, to the ISS-KKLT superpotential\footnote{
For $w_0 \neq 0$, we cannot make the vevs of fields real without 
a fine-tuning of complex parameters, because in such case the 
number of complex parameters is four.}, 
\begin{eqnarray}
W &=& w_0-Ae^{-aT} +Ce^{-cT}+Be^{-bT}X.
\label{eq:issrt}
\end{eqnarray}
We call this ISS-racetrack model. 
In this case, we find $V_T|_0 \sim b e^G|_0$ and\footnote{
This argument depends on parameters in the superpotential. 
For example, when the vacuum of modulus is almost supersymmetric 
Minkowski vacuum, that is, 
$w_0 \sim Ae^{-a\langle T\rangle } 
-Ce^{-c\langle T \rangle } \gg \langle W \rangle$, 
the ISS vacuum is more stable than the usual racetrack vacuum, because 
$m_T$ is much heavier than  $ac(T+\bar{T})^2m_{3/2}$ for such case.} 
\begin{eqnarray}
\hat{V}_{IJ} \big|_0 
\sim \left( 
\begin{array}{cc}
1/X_0 & abc \\
abc &  a^2c^2
\end{array} \right)\,e^G|_0. 
\nonumber
\end{eqnarray}
Note that the modulus mass is estimated as 
\begin{eqnarray}
m_T &\sim& ac(T_0+\bar{T_0})^2m_{3/2}, 
\label{eq:mtissrt}
\end{eqnarray}
without a fine-tuning among $a,~b$ and $c$.
Therefore, we typically find
\begin{eqnarray}
\frac{\delta T}{T_0} \sim \frac{b}{a^2c^2},~~~\frac{\delta X}{X_0}
\sim \frac{b^2}{ac},
\nonumber
\end{eqnarray}
and the reference point is stable so far as $b^2 < ac$. 
Here, we consider the case that $b \lesssim c \lesssim a$.
The K\"ahler potential mixing can be safely 
introduced into the ISS-racetrack model without affecting the structure 
of the minimum due to the smallness of $X_0$. 
Finally, in this ISS-racetrack model, the vacuum would be metastable. 
However its life time would be sufficiently 
long~\cite{Intriligator:2006dd,Dudas:2006gr,Westphal:2007xd}, because 
$Be^{-b \langle T \rangle} \ll 1$.

\section{SUSY breaking order parameters}
\label{sec:orderparameter}
In the previous section, we have studied generalized F-term uplifting 
scenarios where 
the hidden sector field $X$, which is responsible for the nonvanishing 
F-term, is directly couples to the light K\"ahler modulus $T$,  
e.g., because of stringy instanton effects~\cite{Florea:2006si}\footnote{ 
As shown in~\cite{Acharya:2007rc}, in such a case we may have 
a nice property that the smallness 
of the gravitino mass $m_{3/2} \ll M_{Pl}$ could be a natural 
consequence of a tiny cosmological constant.}. 
As we can see from Eqs.~(\ref{eq:mtpkklt}) and (\ref{eq:mtissrt}), 
the $X$-$T$ mixing generically produce a larger hierarchy 
$m_T/m_{3/2} \sim a^2$ between the gravitino mass and the modulus 
mass than one in the original KKLT model $m_T/m_{3/2} \sim a$, 
because the modulus superpotential is effectively given by the 
racetrack-type~\cite{Choi:2005ge}. 
Otherwise, the reference point is unstable and the KKLT-type 
structure that the modulus is somehow heavier than 
the gravitino, might be spoiled as shown in the ISS-KKLT model. 

In this section, we analyze the SUSY breaking order parameters in detail.
We will find that, despite the large modulus mass $m_T \sim a^2 m_{3/2}$, 
we can still obtain the same size of the modulus F-component  
$F^T \sim m_{3/2}/a$ as one in the original KKLT scenario, 
because of the 
enhancement by the $X$-$T$ mixing in the superpotential.  
In the following, we first derive general expressions for the 
order parameters $F^T$ and $F^X$ in terms of modulus and gravitino 
masses, and then apply them to several typical models.

\subsection{General result}
\label{ssec:gen}
For the evaluation of $F^T$, we expand the relevant part of 
the scalar potential around the reference point defined by 
(\ref{eq:refpt}) with $W|_0 \neq 0$. 
\begin{eqnarray}
V 
&=& e^K (K^{I\bar{J}}D_IW\overline{D_{{J}}W}-3|W|^2) 
\nonumber \\
&\approx & V \big|_0 
+[e^K K^{X\bar{X}}\overline{D_XW}
(K_{XT}W+K_XW_T +W_{XT})] \big|_0 \delta T 
\nonumber \\ && 
+[e^K K^{T\bar{T}}(K_{TT}W+K_TW_T+W_{TT})] \big|_0 
\overline{D_TW}\,\delta T 
-3[e^K\bar{W}W_T] \big|_0 \delta T + \cdots 
\nonumber \\
&\approx&  [e^K K^{X\bar{X}}\overline{D_XW}
(-K_XK_TW +W_{XT})] \big|_0 \delta T 
\nonumber \\ && 
+[e^K K^{T\bar{T}}W_{TT}] \big|_0 \overline{D_TW}\,\delta T
+3[K_Te^K|W|^2] \big|_0 \delta T 
+\cdots ,
\nonumber
\end{eqnarray}
where ellipses denote complex conjugates and terms that are of 
order $(\delta X)^2$, $(\delta \bar{X})^2$, $(\delta T)^2$, 
$(\delta \bar{T})^2$ and $|\delta X|^2$. Here we have assumed
\begin{eqnarray}
& & 
|K_{XT}|, \ |K_{X\bar{T}}| 
\ < \ 
\left| \frac{W_{XT}}{W_{TT}} \right| 
\ \sim \ {\cal O}(a^{-1})
\ < \ 
K_{T\bar{T}}, \ K_{X\bar{X}} 
\ \sim \ 1 ,
\nonumber \\
& & |K_{TT}W|, \ |K_{T}W_T| = |-K_T^2W| 
\ \ll \ |W_{TT}|, 
\nonumber \\
& & |V_{TT}| 
\ \ll \ |V_{T\bar{T}}|, 
\nonumber
\end{eqnarray}
at the reference point. 
An important point is that the expansions of $\overline{D_TW}$ are 
all sub-leading, so the order parameter $F^T$ can be mainly of 
${\mathcal O}(\delta \bar{T},~\delta \bar{X})$. As a result, 
terms which are of 
${\mathcal O}(\delta T, |\delta T|^2,~\delta T \delta \bar{X})$ 
in the scalar potential are important for our analysis. 

Using the expressions $m_T \simeq -e^{K/2}K^{T\bar{T}}W_{TT} \Big|_0$ 
and $F^T \simeq -e^{K/2}K^{T\bar{T}} \Big|_0 \overline{D_{T}W} $, 
one can rewrite the above scalar potential 
as
\begin{eqnarray}
V &\approx& \delta T\left[ m_{3/2}^2 
\left\{ \sqrt{3K^{X\bar{X}}}\left(\frac{W_{TX}}{W}-K_XK_T \right)
+3K_T\right \} 
+K_{T\bar{T}}m_T F^T \right] + \cdots,
\nonumber
\end{eqnarray}
where we have omitted the symbols $\big|_0$ and 
we have used $\sqrt{K^{X\bar{X}}}D_XW = \sqrt{3}W$.
With a equation of motion for $\delta T$, we obtain
\begin{eqnarray}
F^T &\simeq& \frac{m_{3/2}^2}{m_T}
\left( \frac{-K_T}{K_{T\bar{T}}} \right) 
\left[ \sqrt{3}(\sqrt{3}-\sqrt{K^{X\bar{X}}}K_X)  
+ \sqrt{3K^{X\bar{X}}}\left(\frac{W_{TX}}{K_TW}\right) \right]
\nonumber \\
&=& \frac{m_{3/2}^2}{m_T}
\left( \frac{-K_T}{K_{T\bar{T}}} \right)
\sqrt{3}(\sqrt{3}-\sqrt{K^{X\bar{X}}}K_X) 
\left( 1+\frac{\partial_T\ln(W_X)}{K_T} \right),
\label{eq:ftgen}
\end{eqnarray}
where the effect of the $X$-$T$ mixing is encoded in 
$\partial_T \ln W_X=b$ as well as $m_T$. 
Here we have used $\sqrt{K^{X\bar{X}}}D_XW = \sqrt{3}W$ and 
$$
\frac{W_{TX}}{W} 
\ = \ 
\frac{W_{TX}}{W_X}\cdot \frac{W_{X}}{W} 
\ = \ 
\frac{W_{TX}}{W_X}\cdot 
\frac{(\sqrt{3}- \sqrt{K^{X\bar{X}}} K_X )}{
\sqrt{K^{X\bar{X}}}},$$
at the reference point. 
Form Eq.~(\ref{eq:ftgen}), we easily find that $F^T \sim m_{3/2}/a$ 
for the typical modulus mass $m_T/m_{3/2} \sim ab$ shown in 
Eqs.~(\ref{eq:mtpkklt}) and (\ref{eq:mtissrt}). This should be compared 
with the original KKLT model and the F-term uplifting scenarios without 
the mixing $\partial_T \ln W_X=0$, which result in 
$F^T \sim m_{3/2}/a$ with $m_T/m_{3/2} \sim a$. 
The ratio $F^T/m_{3/2}$ can be of the same order as the one 
in the original KKLT model $F^T/m_{3/2} \sim 1/a$, although 
our models have a larger 
hierarchy between the modulus and the gravitino masses. 
This is due to the enhancement factor $\partial_T \ln W_X=b$ in 
Eq.~(\ref{eq:ftgen}).  
On the other hand, around the reference point, we simply find 
$$F^X  \simeq  -\sqrt{3/K_{X\bar{X}}}\,m_{3/2}.$$ 

Here, we comment on the shift of field vevs from the reference point. 
{}From the above expressions, we can roughly estimate as 
$F^T \sim \delta T \partial_{\bar{T}}F^T \sim m_T \delta T 
\sim b m_{3/2}^2/m_T$ where $b= -\partial_T\ln(W_X)$. 
Then we find 
\begin{eqnarray}
\frac{\delta T}{T_0} 
&\sim& b \frac{m_{3/2}^2}{m_T^2},
\nonumber
\end{eqnarray}
where we have assumed $T_0={\cal O}(1)$. 
Furthermore, $\delta X$ in later examples can be typically given by 
$\delta X 
\sim -\delta T (V_{X\bar{T}}/V_{X\bar{X}}) 
\sim -\delta T \frac{bm_T m_{3/2}}{m_{3/2}^2(1+b^2X_0)/X_0}$.
Then we find 
\begin{eqnarray}
\frac{\delta X}{X_0} 
&\sim& -\frac{b^2m_{3/2}}{m_{T}(1+b^2X_0)}. 
\nonumber
\end{eqnarray}
These expressions agree with results in the previous sections 
and are useful for checking the stability of the reference point 
in each model. 

In the following subsections, we show several concrete examples. 
In some cases, we will assume that the gauge kinetic function of 
the hidden sector $f_{hid}$, which is responsible for the 
nonperturbative superpotential terms, is given by the mixture of 
the heavy and the light moduli~\cite{Abe:2005rx}, e.g., 
$f_{hid}=wT+m \langle S \rangle$, for generality. 
All the parameters in the models 
are taken to be positive and real\footnote{Except for model 5, 
we can always assume the positive and real parameters up to 
the overall phase of the superpotential. In model 5, we have to 
tune some of these complex parameters for such assumption.}, and 
we will take $b < a$ in the Polonyi-KKLT model and $b \leq c<a$ 
in the ISS-racetrack model for concreteness. 


\subsection{Polonyi-KKLT model}
\label{ssec:ex:pkklt}
First examples are some simplified versions of the Polonyi-KKLT model 
(\ref{eq:super:pkklt}). As we emphasized, the perturbation around 
the Polonyi vacuum $X_0 \sim {\cal O}(1)$ generates effectively 
a racetrack-type superpotential for the light modulus $T$. Then, we can simply 
drop the constant piece $w_0$ in the superpotential (\ref{eq:super:pkklt}) 
just for stabilizing $T$ unlike the case without the $X$-$T$ mixing. 
Instead, we introduce a moduli mixing in the hidden sector gauge kinetic 
function~\cite{Abe:2005rx}.

\subsubsection{Model 1}

We start from the K\"ahler potential and the 
superpotential given by 
\begin{eqnarray}
K &=& -3\ln(T+\bar{T})+|X|^2, \qquad 
W \ = \ Ae^{-c\langle S \rangle + aT} +Be^{-bT}X, 
\nonumber
\end{eqnarray}
where the dilaton $S$ is assumed to be stabilized at a much higher 
scale by, e.g., the flux induced superpotential. In this case, we find 
\begin{eqnarray}
X_0 &=& \sqrt{3}-1, \qquad 
T_0 \ \simeq \ 
\frac{1}{a+b} \ln \left[ 
\frac{bBX_0}{aA}e^{c \langle S \rangle} \right]
\ \sim \ 
\frac{c}{a+b}\langle S \rangle, 
\nonumber \\
\langle W_{TT} \rangle 
&\simeq & 
ab \langle W \rangle, \qquad 
\langle W \rangle 
\ \simeq \ 
Ae^{-c  \langle S \rangle}
\left( \frac{bBX_0e^{c  \langle S \rangle}}{aA} \right)^{\frac{a}{a+b}} 
\left( 1+\frac{a}{b} \right), 
\nonumber \\
F^T & \simeq & 
-\sqrt{3}\frac{m_{3/2}}{a}, \qquad 
m_{T} \ = \ -\frac{ab(T+\bar{T})^2}{3}m_{3/2}. 
\nonumber
\end{eqnarray}
Therefore the condition for the Polonyi vacuum, $W_X \big|_0=W \big|_0$, 
i.e. $V\big|_0=0$ at $X_0 = \sqrt{3}-1$,  
requires $b/a \simeq 1/X_0 -1\simeq 0.37$, and the positive sign $+aT$ 
in the first exponent is necessary in the superpotential 
in order for the 
Polonyi-like vacuum to be compatible with the racetrack vacuum\footnote{
In the ISS-KKLT model with $W = Ae^{\mp aT}+Be^{\pm bT}X$, we cannot satisfy 
this kind of condition $W_X \big|_0 \simeq \sqrt{3} W \big|_0$, that is, 
$b/a \simeq 1/(\sqrt{3}X_0)-1 \sim 10^{12}$.}.

To evaluate the ratio of the anomaly mediation to the modulus
mediation, we define 
$\alpha$ as~\cite{Choi:2005uz} 
\begin{eqnarray}
\alpha &\equiv& 
\frac{1}{\ln(M_{Pl}/m_{3/2})} \cdot 
\frac{F^C/C_0}{{F^T}/({T+\bar{T}})}, 
\label{eq:alpha}
\end{eqnarray}
where $C_0 = e^{K/6}$ is the lowest component of the conformal
compensator  superfield $C$, and $F^C$ is F-component of $C$, i.e.
\begin{eqnarray}
\frac{F^C}{C_0} = m_{3/2} + \frac{1}{3}\sum_I K_IF^I.
\end{eqnarray}
Note that $F^C/C_0$, in general, has contributions due to 
the gravitino mass and F-components $F^I$ of the SUSY breaking sector, 
although we have $F^C/C_0=m_{3/2}$ in the original KKLT scenario. 
We obtain in this example 
\begin{eqnarray}
\alpha 
&=& -\frac{2}{3} 
 \frac{a}{b} 
\ \simeq \ 
-\frac{2}{3} (\sqrt{3}+1) 
\ \simeq \ -1.82. 
\nonumber
\end{eqnarray}
where 
$\ln(M_p/m_{3/2}) \simeq \ln(1/W \big|_0) \simeq bT_0$ 
has been adopted\footnote{Contributions to 
$F^C/C_0 = m_{3/2}/\sqrt{3} $ 
come from the gravitino mass and 
$K_XF^X/3 = -(1-1/\sqrt{3}) m_{3/2}$ 
in the Polonyi-KKLT model.
The latter  was not taken into account  
in~\cite{Abe:2006xp}, and the value of $\alpha$ was evaluated 
for $F^C/C_0 = m_{3/2}$, but that should be replaced by 
$F^C/C_0 = m_{3/2}/\sqrt{3} $.}. 

\subsubsection{Model 2}

We can also consider the case with a moduli mixing in the 
term which is responsible for the $X$-$T$ 
mixing through, e.g., stringy instanton effects. 
Then we analyze the following model: 
\begin{eqnarray}
K &=& -3\ln(T+\bar{T})+|X|^2, \qquad 
W \ = \ Ae^{- aT} +Be^{-c \langle S \rangle + bT}X. 
\nonumber
\end{eqnarray}
Note that the positive sign $+bT$ in the second exponent 
is necessary in order for the Polonyi-like vacuum to be compatible 
with the racetrack vacuum as in the previous model. 
In this case, we obtain 
\begin{eqnarray}
X_0 &=& \sqrt{3}-1, \qquad 
T_0 \ \simeq \ 
\frac{1}{a+b}\ln \left[\frac{aA}{bBX_0} 
e^{c \langle S \rangle} \right]
\ \sim \ 
\frac{c}{a+b}\langle S \rangle, 
\nonumber \\
\langle W_{TT} \rangle 
&\simeq & ab \langle W \rangle, \qquad 
\langle W \rangle 
\ \simeq \ 
A \left( \frac{aAe^{c\langle S \rangle}}{bBX_0} 
\right)^{-\frac{a}{a+b}}\left(1+\frac{a}{b} \right), 
\nonumber \\
F^T &\simeq& \sqrt{3}\frac{m_{3/2}}{a}, \qquad 
m_{T}=-\frac{ab(T+\bar{T})^2}{3}m_{3/2}.
\nonumber
\end{eqnarray}
The condition for realizing the Polonyi vacuum is the same 
as the previous example, $b/a \simeq 1/X_0-1$. 
The anomaly/modulus mediation ratio (\ref{eq:alpha}) is found in this model as 
\begin{eqnarray}
\alpha &=& \frac{2}{3}
\frac{1}{-\frac{b}{a}+\frac{c \langle S \rangle}{aT_0}} 
\ = \ 
\frac{2}{3} 
\nonumber
\end{eqnarray}
where we applied 
$\ln(M_p/m_{3/2}) = c\langle S \rangle -bT_0 \simeq aT_0$. 

\subsubsection{Model 3}
Finally, we show the results in the Polonyi-KKLT model 
(\ref{eq:super:pkklt}) with the minimal K\"ahler potential: 
\begin{eqnarray}
K &=& -3\ln(T+\bar{T})+|X|^2, \qquad 
W \ = \ w_0-Ae^{-aT} +Be^{-bT}X, 
\nonumber
\end{eqnarray}
where we find 
\begin{eqnarray}
X_0 &=& \sqrt{3}-1 , \qquad 
T_0 \ \simeq \ \frac{1}{a-b}\ln \left[\frac{aA}{bBX_0} \right], 
\nonumber \\
\langle W_{TT} \rangle 
&\simeq & -ab X_0 \left(1-\frac{b}{a} \right)\langle W \rangle, \qquad 
\langle W \rangle 
\ \simeq \ 
B\left(\frac{aA}{bBX_0}\right)^{-\frac{b}{a-b}}, 
\nonumber \\
F^T &\simeq& 
\sqrt{3}\frac{m_{3/2}}{(a-b)X_0}, \qquad 
m_{T} \ = \ 
\frac{ab(T+\bar{T})^2}{3}\left( 1- \frac{b}{a}\right)X_0m_{3/2}.
\nonumber
\end{eqnarray}
Here we adopted 
$w_0 \simeq Ae^{-aT_0}+B(1-X_0 )e^{-bT_0}$ 
coming from the condition for the Polonyi-like vacuum, 
$W \big|_0 = W_X \big|_0= Be^{-bT_0}$. 

In this case, one finds that the shift of fields are given by 
$\delta T/T_0 \sim 1/b(a-b)^2$ and $\delta X/X_0 \sim 1/b(a-b)$. 
Then the expansion of the potential around the reference point 
is done in terms of $1/(a-b)^2$, where 
\begin{eqnarray}
D_XW &=& D_XW \big|_0 
+ \sum_{n=1}\frac{1}{n!}(\partial_T^n W_{X}) \big|_0 (\delta T)^n+ \cdots 
\nonumber \\ &\sim& 
m_{3/2}\left(1 + \frac{1}{(a-b)^2} + \frac{1}{(a-b)^4}  
+\cdots \right), 
\nonumber \\
D_TW &=& 
\sum_{n=1}\frac{1}{n!}(\partial_T^n W_{T}) \big|_0(\delta T)^n 
+\delta X \sum_{n=1}\frac{1}{(n-1)!}(\partial_{T}^n W_{X}) \big|_0 
(\delta T)^{n-1}+ \cdots 
\nonumber \\ &\sim& 
\frac{m_{3/2}}{(a-b)}\left( 1+  \frac{1}{(a-b)^2} 
+\frac{1}{(a-b)^4}+ \cdots \right). 
\nonumber
\end{eqnarray}
Therefore it is a good approximation when $a-b={\cal O}(10)$. 
Note that only if $B=e^{-c\langle S \rangle}$ with $c \sim a \sim b$, 
the difference $a-b$ can be of ${\cal O}(10)$ for 
$T_0 \sim \langle S \rangle = {\mathcal O}(1)$. 

In such a case, the mirage mediation is important 
for the gaugino masses. For $c \neq 0 $ we find the 
anomaly/modulus mediation ratio (\ref{eq:alpha}) as 
\begin{eqnarray}
\alpha &=& 
\frac{2}{3} 
\frac{(a-b)X_0}{\left(b+c\frac{\langle S \rangle}{T_0} \right)} 
\ = \ 
\frac{2}{3} (\sqrt{3}-1)
{\left(1-\frac{b}{a} \right)} 
\ < \ 0.49, 
\nonumber
\end{eqnarray}
where we substituted 
$\ln(M_p/m_{3/2}) \simeq  c\langle S \rangle +bT_0 \simeq aT_0$. 

For $c=0$, $F^T$ is of ${\mathcal O}(m_{3/2})$ through 
a tuning $(a-b)^2 = {\mathcal O}(1)$ in order to make $T_0$ 
sufficiently large. However, in this case the perturbation around 
the reference point becomes unstable. 
Thus,  generically it is required that  
$\delta T/T_0 \leq 1/b^2$ and $\delta X/X_0 \leq 1/b$ 
for the convergence.

\subsection{ISS-racetrack model}
\label{ssec:ex:issrt}

\subsubsection{Model 4}
Next we analyze the ISS-racetrack model~(\ref{eq:issrt}). 
Similarly we first omit the constant piece $w_0$ 
in the superpotential and start with 
\begin{eqnarray}
K &=& -3\ln(T+\bar{T})+|X|^2-\frac{|X|^4}{\Lambda^2}, \qquad 
W \ = \ -Ae^{-aT} +Ce^{-cT}+Be^{-bT}X. 
\nonumber
\end{eqnarray}
Then we find the following results.
\begin{eqnarray}
X_0 &=& \frac{\sqrt{3}}{6}\Lambda^2 
\ \sim \ 
16\pi^2 B e^{-bT_0} \simeq 10^2m_{3/2}
\ \ll \ 1, \qquad 
T_0 \ \simeq \ 
\frac{1}{a-c}\ln \left[\frac{aA}{cC} \right], 
\nonumber \\
\langle W_{TT} \rangle 
&\simeq & -ac \langle W \rangle, \qquad 
\langle W \rangle 
\ \simeq \ 
\frac{B}{\sqrt{3}}\left(\frac{aA}{cC}\right)^{-\frac{b}{a-c}} 
\ \simeq \ 
A \left ( \frac{a}{c}-1\right ) \left( \frac{aA}{cC} 
\right)^{-\frac{a}{a-c}}, 
\nonumber \\
F^T & \simeq & 3 \frac{b}{ac}m_{3/2}, \qquad 
m_{T} \ = \ \frac{ac(T+\bar{T})^2}{3}m_{3/2}, 
\nonumber
\end{eqnarray}
where the condition for obtaining the ISS-like vacuum 
$W \big|_0 \simeq D_XW \big|_0/\sqrt{3} \simeq 
W_X \big|_0/\sqrt{3}$, i.e. $V\big|_0 =0$, has been adopted. 
The anomaly-to-modulus mediation ratio (\ref{eq:alpha}) 
is now given by 
\begin{eqnarray}
\alpha &=& \frac{2acT_0}{3b\ln(M_p/m_{3/2})} 
\ = \ \frac{2ac}{3b^2}, 
\nonumber
\end{eqnarray}
where we applied $\ln(M_p/m_{3/2}) \simeq bT_0$. 
Note that $ac/b^2>1$ is required in order to 
make the reference point of the ISS model stable. 
For the case with $ac/b^2=3$, we obtain $\alpha=2$.
For the case with $b=c,~B=C=O(1)$, we need a large value of $A$, 
e.g., $A=e^{-d\langle S\rangle} \gg 1$ and the condition that $b/a \simeq 1-1/\sqrt{3} \simeq 0.42$
to make the reference point stable, and then find $\alpha \simeq 1.58$.

\subsubsection{Model 5}
Finally we turn on the constant $w_0$ in the previous example: 
\begin{eqnarray}
K &=& -3\ln(T+\bar{T})+|X|^2-\frac{|X|^4}{\Lambda^2}, \qquad 
W \ = \ w_0-Ae^{-aT} +Ce^{-cT}+Be^{-bT}X. 
\nonumber
\end{eqnarray}
Then we obtain the following results, 
\begin{eqnarray}
X_0 &=& \frac{\sqrt{3}}{6}\Lambda^2 
\ \simeq \ 10^2m_{3/2} 
\ \ll \ 1, \qquad 
T_0 \ \simeq \ 
\frac{1}{a-c}\ln \left[\frac{aA}{cC} \right], 
\nonumber \\
\langle W_{TT} \rangle &\simeq& 
-ac \left( \frac{a}{c} -1 \right)A e^{-aT_0}, \qquad 
\langle W \rangle 
\ \simeq \  
\frac{B}{\sqrt{3}}e^{-bT_0}
\ \simeq \ 
\frac{B}{\sqrt{3}}\left(\frac{aA}{cC}\right)^{-\frac{b}{a-c}}, 
\nonumber \\
{F^T} &\simeq& 
\frac{\sqrt{3}bB}{(a-c)aA}e^{(a-b)T_0}m_{3/2}
\ \simeq \ 
\frac{\sqrt{3}bB}{(a-c)aA} \left( \frac{aA}{cC} 
\right)^{\frac{a-b}{a-c}}m_{3/2}
\nonumber \\
m_{T} &=& 
\frac{ac(T+\bar{T})^2}{3}m_{3/2} 
\left[\frac{\sqrt{3}A}{B}\left( \frac{a}{c}-1\right) 
e^{-(a-b)T_0}\right], 
\nonumber
\end{eqnarray}
where we applied the condition for realizing the ISS-like vacuum, 
$w_0 \simeq (B/\sqrt{3}) e^{-bT_0} +Ae^{-aT_0}-Ce^{-cT_0}$. 

The anomaly/modulus mediation ratio (\ref{eq:alpha}) is found as 
\begin{eqnarray}
\alpha &=& \frac{2(a-c)aA}{\sqrt{3}bB }e^{-(a-b)T_0}
\frac{T_0}{\ln \left(\frac{M_p}{m_{3/2}}\right)}. 
\nonumber
\end{eqnarray}
Note that 
$\ln(M_{Pl}/m_{3/2}) \simeq \ln(B^{-1}e^{bT_0})$. 
We find that $\alpha$ is sensitive to the adjustment of $B$. 
For $B = {\mathcal O}(1)$, we obtain 
$\ln\left(\frac{M_p}{m_{3/2}}\right) \simeq bT_0 $. 
Therefore, 
\begin{eqnarray}
\alpha &=& \frac{2(a-c)aA}{\sqrt{3}b^2B }e^{-(a-b)T_0}. 
\nonumber
\end{eqnarray}
In the case that $A$ and $C$ are of ${\mathcal O}(1)$, 
a value of $a-c$ must be also of ${\mathcal O}(1)$ for large $T_0$. 
Thus, the value of $\alpha$ becomes typically very small such as 
$\alpha \sim b^{-1}e^{-(a-b)T_0}$. However in this case 
it is not valid to use the reference point for the analysis. 
For example, we take the parameters as 
$c=b$ and $C = B={\mathcal O(1)}$ and one finds that 
$m_T \sim a(a-b)m_{3/2}$ and then 
$\delta T/T_0 \sim b/a^2(a-b)^2 \sim 1/b$, 
$\delta X/X_0 \sim b^2/a(a-b) \sim b$. 
Therefore the perturbation of the potential around the reference 
point does not converge. 
On the other hand, if $A=e^{d\langle S \rangle}$ 
with $d = {\mathcal O}(10)$, a value of $a-c$ can be of ${\mathcal O}(10)$. 
For example, in the case with $b=c$ and $B=C$, we find 
$\alpha \simeq (2/\sqrt{3})\left(a/b-1\right) 
={\mathcal O}(1)$ and the reference point is stable. 

For $B=Ce^{-d\langle S \rangle} \ll 1$ 
with $d= {\mathcal O}(1-10) $, 
we obtain $\ln(M_p/m_{3/2}) \simeq d\langle S \rangle + bT_0$ and find 
\begin{eqnarray}
\alpha &=& 
\frac{2(a-c)aA}{\sqrt{3}bC} 
\frac{1}{b+d\frac{\langle S \rangle}{T_0}} 
e^{d\langle S \rangle-(a-b)T_0}. 
\nonumber
\end{eqnarray}
In this case, the magnitude of $\alpha$ can be strongly 
dependent on a $e^{-d\langle S \rangle}$. 
For example, when $A$, $C$, $a-b ={\cal O}(1)$ 
and $b=c$, one obtains 
$F^T=\sqrt{3}e^{-d\langle S \rangle}m_{3/2}/(a-b)$ and
\begin{eqnarray}
\alpha &=& 
\frac{2(a-b)}{\sqrt{3}} 
\frac{1}{b+d\frac{\langle S \rangle}{T_0}}e^{d\langle S \rangle}. 
\nonumber
\end{eqnarray}
For the case with $b \lesssim e^{d\langle S \rangle} $, i.e. 
$d \langle S \rangle ={\cal O}(1)$, 
the value of $\alpha$ becomes of ${\cal O}(1)$ 
and shifts of fields are given by 
$\delta T/T_0 \sim e^{-2d\langle S \rangle }/b(a-b)^2 
\lesssim 1/b^3$ and 
$\delta X/X_0 \sim b e^{-d\langle S \rangle}/(a-b) 
\lesssim 1/(a-b) \lesssim 1$. 

On the other hand, for the case with $b \ll e^{d\langle S \rangle}$, 
i.e. $d \langle S \rangle ={\cal O}(10)$, one can obtain
$\alpha \sim e^{d\langle S \rangle} \gg 1$ and find that the 
anomaly mediation is dominant compared with the modulus mediation. 
In the latter case, note that mass of the modulus can be much heavier 
than the gravitino mass such as 
\begin{eqnarray}
m_{T} &\simeq& 
\frac{(a-b)b(T+\bar{T})^2}{\sqrt{3}}
e^{d\langle S \rangle}m_{3/2} 
\ \sim \ 
e^{d\langle S \rangle}m_{3/2} 
\ \gg \ ab (T+\bar{T})^2m_{3/2},
\nonumber
\end{eqnarray}
and the magnitude of $w_0$ becomes almost $Ae^{-aT_0}-Ce^{-cT_0}$ 
which can be much larger than $Be^{-bT_0}$ 
like the model in~\cite{Kallosh:2004yh,Kallosh:2006dv}. 
In this case, the vacuum of the modulus can be stable during the 
inflation and then the modulus-induced gravitino problem can be avoided. 
The Polonyi problem may not occur if we have a low scale inflation like 
a new inflation. In that case, if one changes the gravitino mass of 
${\cal O}(100)$ TeV to ${\cal O}(1)$ GeV, one could 
realize the gauge mediation model, 
which is studied in~\cite{Kitano:2006wz,de Alwis:2007qx}. 
However that is beyond the scope of this paper.

\subsection{Phenomenological aspects}
\label{ssec:pheno}
We have studied several concrete models.
Here we comment on their phenomenological aspects like 
SUSY spectra in the visible sector.
In most of models except model 5, the modulus mass $m_T$ is 
quite large compared with the gravitino mass $m_{3/2}$, 
i.e. 
\begin{eqnarray}
\frac{m_T}{m_{3/2}}={\cal O}(a^2), 
\nonumber 
\end{eqnarray}
with $a \sim 4 \pi^2$.
Model 5 can derive much heavier modulus mass.

There are three important sources of SUSY breaking, $F^X$, $F^T$ and 
$F^C$, where SUSY breaking through $F^C$ appears as 
the anomaly meditation.
Most of models except model 5 predict similar ratios among 
$F^X$, $F^T$ and $F^C$.
Thus, first we concentrate ourselves to models 1-4.
These models have $F^X={\cal O}(m_{3/2})$ and 
\begin{eqnarray}
\frac{F^T}{m_{3/2}} = {\cal O}(a^{-1}). 
\nonumber
\end{eqnarray}
The ratio of the anomaly mediation to the modulus 
mediation is denoted by a value of $\alpha$ 
defined in Eq.~(\ref{eq:alpha}). 
Models 1-4 lead to $\alpha = {\cal O}(1)$, 
that is, the modulus mediation and anomaly mediation 
are comparable.

Now, let us evaluate soft SUSY breaking terms 
in the visible sector.
For such purpose, we have to fix couplings between 
SUSY breaking sources and the visible sector.
We assume that gauge kinetic functions of the visible sector 
$f_v$ are obtained as 
\begin{eqnarray}
f_v = w_vT+m_vS, 
\nonumber
\end{eqnarray}
where $w_v$ and $m_v$ are constants.
In the simplest case with $w_v=1$ and $m_v=0$, 
gaugino masses in the visible sector appear as 
the mirage mediation with the values of $\alpha$, 
which are shown in the previous subsections.
In generic case, gaugino masses are obtained 
as the mirage mediation by replacing $\alpha$ by 
$\alpha \frac{f_v + \bar f_v}{w_v(T +\bar T)}$.
For example, we could derive the value of 
$\alpha \approx 2$ 
in the simplest case with $w_v=1$ and $m_v=0$, 
e.g. in model 4 with $ac/b^2=3$.
Also, other cases with nonvanishing values of 
$w_v$ and/or $m_v$ would lead to 
$\alpha \frac{f_v + \bar f_v}{w_v(T +\bar T)} \approx 2$.
In these models, gaugino masses at the TeV scale are given 
by the (tree level) pure modulus mediation, that is, 
the TeV scale mirage unification of gaugino 
masses~\cite{Choi:2005uz, Choi:2005hd}.
If the $X$ field couples to heavy modes, which are 
charged under the SM gauge group, effects due to $F^X$ 
could appear through loop-effects, that is, the 
gauge mediation.
Its contribution is the same size as the anomaly mediation
when mass of the messenger fields is Planck scale.
Also, if  the $X$ field does not sequestered from 
visible matter fields in the K\"ahler metric and the VEV 
of the $X$ field is comparable with the Planck scale, 
effects due to $F^X$ could appear in the gaugino mass 
through the Konishi/K\"ahler anomaly~\cite{Bagger:1999rd}.\footnote{
See also~\cite{Choi:2007ka}.}  
At any rate, the gaugino masses are of ${\cal O}(m_{3/2}/a)\sim 
1$ TeV unless a direct coupling appears in the 
gauge kinetic function.

Now, we also consider the couplings between 
the SM matter fields and $X$.
In our framework, the hidden sector field $X$ directly 
couples to the modulus $T$.
Thus, it would be natural that $X$ might live in the Calabi-Yau space 
rather than in the warped throat.  
Also the SM lives on D-branes which are wrapping on bulk 
Calabi-Yau.
Thus, the contact terms between $X$ and 
the SM matter fields $Q$, 
\begin{eqnarray}
\int d \theta^4 c|X|^2|Q|^2, 
\nonumber
\end{eqnarray}
would not be suppressed.
Then, we would obtain SUSY braking scalar mass and 
A-terms of ${\cal O}(m_{3/2}) \sim 100$ TeV.
In this case, we have a large hierarchy between 
gaugino masses $M_a$ and scalar masses $m_i$ as 
\begin{eqnarray}
M_a={\cal O}(m_{3/2}/a), \qquad m_i={\cal O}(m_{3/2}),
\nonumber
\end{eqnarray}
with $a={\cal O}(4 \pi^2)$.
That would have several phenomenological interesting 
aspects~\cite{Wells:2004di}.
If the SM matter fields are sequestered from $X$ 
by any reason and the above contact terms are 
suppressed sufficiently, 
the mirage mediation would also be dominant 
in scalar masses and A-terms.

The situation in model 5 is different from others.
Model 5 has a rich structure in the modulus mass and 
ratios among $F^X$, $F^T$ and $F^C$.
A heavier modulus mass could be realized in model 5, 
and a value of $\alpha$ can vary from values 
like $\alpha ={\cal O}(1)$ to large values 
like $\alpha ={\cal O}(10)$.
In the latter case with  $\alpha ={\cal O}(10)$, 
the anomaly mediation would be dominant in 
gaugino masses in the visible sector.
Scalar masses would be of ${\cal O}(m_{3/2})$ 
unless the SM matter fields are sequestered from $X$ 
by any reason.

Finally, we comment on the phases of F-components. 
The phases are aligned as
${\rm Arg}[F^C] = {\rm Arg}[F^X] ={\rm Arg}[F^T]={\rm Arg}[\bar{W}] $
as long as ${\rm Arg}[W_{TT}] ={\rm Arg}[W] $.
Thus, we can always solve the SUSY CP problem except for model 5.
In model 5, we need a find-tuning for the alignment unless $\alpha =O(10)$.

\section{Conclusion and discussion}
\label{sec:conclusion}
We have studied modulus stabilization and 
realization of de Sitter vacua through 
F-term uplifting by a dynamically generated F-term. Here the uplifting 
sector $X$ directly couples to the light K\"ahler modulus $T$ in the 
superpotential through, e.g., a stringy instanton effects. 
In the Polonyi-KKLT model, the perturbation from the reference point, 
where the KKLT-type modulus properties and the SUSY breaking structure 
are realized, is stable under the existence of such superpotential 
mixing, though the K\"ahler mixing can spoil the stability in general. 
Contrary, in the ISS-KKLT model, the superpotential mixing makes the 
perturbation unstable, that is, the true minimum is far from the 
reference point. This instability can be avoided by introducing 
another nonperturbative effect into the ISS-KKLT superpotential, i.e., 
by considering the ISS-racetrack model. 

In the case that the perturbation from the reference point is 
stable, the qualitative features of the KKLT scenario are 
preserved even with such modulus mixing to the uplifting sector. 
One of the quantitative changes, which are phenomenologically 
relevant, is a larger hierarchy 
between the modulus mass $m_T$ and the gravitino mass $m_{3/2}$, 
i.e. $m_T/m_{3/2} \sim {\cal O}(a^2)$ . 
Even with such a large mass, the modulus F-term keeps a moderate 
value, $F^T \sim {\cal O}(m_{3/2}/a)$ for $\ln (M_{Pl}/m_{3/2}) \sim a$, 
thanks to the enhancement factor originating from the $X$-$T$ 
mixing in the superpotential. Then we typically find a 
mirage-mediation pattern of gaugino masses of ${\cal O}(m_{3/2}/a)$. 
The scalar masses and the A-terms are generically of ${\cal O}(m_{3/2})$ 
because of possible direct couplings with $X$. 

In the ISS-racetrack model we can realize $m_T \sim 10^{8}$ GeV 
and $m_X \sim 10^{10}$ GeV. In such case we could avoid 
the gravitino overproduction problem~\cite{Nakamura:2006uc} by 
these scalar fields when the inflation has a low Hubble parameter 
$H_{inf} \sim 10^{7}$ GeV. This may be also important for the 
explanation of the dark matter abundance, and in 
this scenario the dark matter candidate is likely to be 
gaugino. Stability of the vacuum during the inflation epoch may 
suppress the non-thermal production of the gaugino 
(neutralino\footnote{
As for the thermal production in 
the mirage-type models, see Refs.~\cite{Baer:2006id}.}) 
and the gravitino through the modulus and $X$ decays. 
With a tuning of $B$, we can partially obtain a sparticle spectrum 
of the anomaly-mediation type, and both the modulus $T$, whose mass can 
be much heavier than $10^{8}$ GeV, and the SUSY breaking field $X$ may 
be stable during the low scale inflation such as a new inflation model. 

\subsection*{Acknowledgement}
The author would like to thank Kiwoon Choi for useful comments.
H.~A.\/,  T.~H.\/ and T.~K.\/ are supported in part by the
Grand-in-Aid for Scientific Research \#182496, \#194494
and  \#17540251, respectively.
T.~K.\/ is also supported in part by 
the Grant-in-Aid for
the 21st Century COE ``The Center for Diversity and
Universality in Physics'' from the Ministry of Education, Culture,
Sports, Science and Technology of Japan.


\begin{thebibliography}{99}


\bibitem{Green:1987sp}
  M.~B.~Green, J.~H.~Schwarz and E.~Witten,
{\it  Cambridge, Uk: Univ. Pr. (1987) 469 P. ( Cambridge Monographs On Mathematical Physics)};
{\it  Cambridge, Uk: Univ. Pr. ( 1987) 596 P. ( Cambridge Monographs On Mathematical Physics)};
  J.~Polchinski,
{\it  Cambridge, UK: Univ. Pr. (1998) 402 p};
{\it  Cambridge, UK: Univ. Pr. (1998) 531 p}

\bibitem{Dine:1985rz}
  M.~Dine, R.~Rohm, N.~Seiberg and E.~Witten,
  Phys.\ Lett.\  B {\bf 156}, 55 (1985).

\bibitem{Nilles:2004zg}
  H.~P.~Nilles,
  arXiv:hep-th/0402022.




\bibitem{Giddings:2001yu}
  S.~B.~Giddings, S.~Kachru and J.~Polchinski,
  Phys.\ Rev.\  D {\bf 66}, 106006 (2002)
  [arXiv:hep-th/0105097];
  S.~Kachru, M.~B.~Schulz and S.~Trivedi,
  JHEP {\bf 0310}, 007 (2003)
  [arXiv:hep-th/0201028];
  A.~Giryavets, S.~Kachru, P.~K.~Tripathy and S.~P.~Trivedi,
  JHEP {\bf 0404}, 003 (2004)
  [arXiv:hep-th/0312104];
  P.~K.~Tripathy and S.~P.~Trivedi,
  JHEP {\bf 0303}, 028 (2003)
  [arXiv:hep-th/0301139];
  O.~DeWolfe and S.~B.~Giddings,
  Phys.\ Rev.\  D {\bf 67}, 066008 (2003)
  [arXiv:hep-th/0208123];
  F.~Denef, M.~R.~Douglas, B.~Florea, A.~Grassi and S.~Kachru,
  Adv.\ Theor.\ Math.\ Phys.\  {\bf 9}, 861 (2005)
  [arXiv:hep-th/0503124];
  S.~Kachru and A.~K.~Kashani-Poor,
  JHEP {\bf 0503}, 066 (2005)
  [arXiv:hep-th/0411279];
  O.~DeWolfe, A.~Giryavets, S.~Kachru and W.~Taylor,
  JHEP {\bf 0507}, 066 (2005)
  [arXiv:hep-th/0505160];
  M.~R.~Douglas and S.~Kachru,
  arXiv:hep-th/0610102;
  F.~Denef, M.~R.~Douglas and S.~Kachru,
  arXiv:hep-th/0701050.

\bibitem{Kachru:2003aw}
  S.~Kachru, R.~Kallosh, A.~Linde and S.~P.~Trivedi,
  Phys.\ Rev.\  D {\bf 68}, 046005 (2003)
  [arXiv:hep-th/0301240].



\bibitem{Choi:2005ge}
  K.~Choi, A.~Falkowski, H.~P.~Nilles and M.~Olechowski,
  Nucl.\ Phys.\  B {\bf 718}, 113 (2005)
  [arXiv:hep-th/0503216];
  K.~Choi, A.~Falkowski, H.~P.~Nilles, M.~Olechowski and S.~Pokorski,
  JHEP {\bf 0411}, 076 (2004)
  [arXiv:hep-th/0411066].

\bibitem{Randall:1998uk}
  L.~Randall and R.~Sundrum,
  Nucl.\ Phys.\  B {\bf 557}, 79 (1999)
  [arXiv:hep-th/9810155];
  G.~F.~Giudice, M.~A.~Luty, H.~Murayama and R.~Rattazzi,
  JHEP {\bf 9812}, 027 (1998)
  [arXiv:hep-ph/9810442].



\bibitem{Choi:2005uz}
  K.~Choi, K.~S.~Jeong and K.~i.~Okumura,
  JHEP {\bf 0509}, 039 (2005)
  [arXiv:hep-ph/0504037].

\bibitem{Endo:2005uy}
  M.~Endo, M.~Yamaguchi and K.~Yoshioka,
  Phys.\ Rev.\  D {\bf 72}, 015004 (2005)
  [arXiv:hep-ph/0504036].

\bibitem{Nakamura:2006uc}
  S.~Nakamura and M.~Yamaguchi,
  Phys.\ Lett.\  B {\bf 638}, 389 (2006)
  [arXiv:hep-ph/0602081];
  M.~Endo, K.~Hamaguchi and F.~Takahashi,
  Phys.\ Rev.\ Lett.\  {\bf 96}, 211301 (2006)
  [arXiv:hep-ph/0602061];
  Phys.\ Rev.\  D {\bf 74}, 023531 (2006)
  [arXiv:hep-ph/0605091];
  T.~Asaka, S.~Nakamura and M.~Yamaguchi,
  Phys.\ Rev.\  D {\bf 74}, 023520 (2006)
  [arXiv:hep-ph/0604132].

\bibitem{Choi:2005hd}
  K.~Choi, K.~S.~Jeong, T.~Kobayashi and K.~i.~Okumura,
  Phys.\ Lett.\  B {\bf 633}, 355 (2006)
  [arXiv:hep-ph/0508029]; 
%
  R.~Kitano and Y.~Nomura,
  Phys.\ Lett.\  B {\bf 631}, 58 (2005)
  [arXiv:hep-ph/0509039]; 
%
  R.~Kitano and Y.~Nomura,
  Phys.\ Rev.\  D {\bf 73}, 095004 (2006)
  [arXiv:hep-ph/0602096]; 
%
  K.~Choi, K.~S.~Jeong, T.~Kobayashi and K.~i.~Okumura,
  Phys.\ Rev.\  D {\bf 75}, 095012 (2007)
  [arXiv:hep-ph/0612258].

\bibitem{Abe:2007kf}
  H.~Abe, T.~Kobayashi and Y.~Omura,
  Phys.\ Rev.\  D {\bf 76}, 015002 (2007)
  [arXiv:hep-ph/0703044].

\bibitem{Intriligator:2006dd}
  K.~Intriligator, N.~Seiberg and D.~Shih,
  JHEP {\bf 0604}, 021 (2006)
  [arXiv:hep-th/0602239].

\bibitem{Forste:2006zc}
  S.~Forste,
  Phys.\ Lett.\  B {\bf 642}, 142 (2006)
  [arXiv:hep-th/0608036];
  M.~Dine, J.~L.~Feng and E.~Silverstein,
  Phys.\ Rev.\  D {\bf 74}, 095012 (2006)
  [arXiv:hep-th/0608159].


\bibitem{Franco:2006es}
  S.~Franco and A.~M.~Uranga,
  JHEP {\bf 0606}, 031 (2006)
  [arXiv:hep-th/0604136];
  H.~Ooguri and Y.~Ookouchi,
  Nucl.\ Phys.\  B {\bf 755}, 239 (2006)
  [arXiv:hep-th/0606061];
  Phys.\ Lett.\  B {\bf 641}, 323 (2006)
  [arXiv:hep-th/0607183];
  R.~Kitano, H.~Ooguri and Y.~Ookouchi,
  Phys.\ Rev.\  D {\bf 75}, 045022 (2007)
  [arXiv:hep-ph/0612139];
  V.~Braun, E.~I.~Buchbinder and B.~A.~Ovrut,
  Phys.\ Lett.\  B {\bf 639}, 566 (2006)
  [arXiv:hep-th/0606166];
  JHEP {\bf 0610}, 041 (2006)
  [arXiv:hep-th/0606241];
  S.~Franco, I.~Garcia-Etxebarria and A.~M.~Uranga,
  JHEP {\bf 0701}, 085 (2007)
  [arXiv:hep-th/0607218];
  A.~Amariti, L.~Girardello and A.~Mariotti,
  JHEP {\bf 0612}, 058 (2006)
  [arXiv:hep-th/0608063].
  I.~Bena, E.~Gorbatov, S.~Hellerman, N.~Seiberg and D.~Shih,
  JHEP {\bf 0611}, 088 (2006)
  [arXiv:hep-th/0608157];
  C.~Ahn,
  Class.\ Quant.\ Grav.\  {\bf 24}, 1359 (2007)
  [arXiv:hep-th/0608160];
  Phys.\ Lett.\  B {\bf 647}, 493 (2007)
  [arXiv:hep-th/0610025];
  JHEP {\bf 0705}, 053 (2007)
  [arXiv:hep-th/0701145];
  R.~Argurio, M.~Bertolini, S.~Franco and S.~Kachru,
  JHEP {\bf 0701}, 083 (2007)
  [arXiv:hep-th/0610212];
  JHEP {\bf 0706}, 017 (2007)
  [arXiv:hep-th/0703236];
  M.~Aganagic, C.~Beem, J.~Seo and C.~Vafa,
  arXiv:hep-th/0610249;
  J.~J.~Heckman, J.~Seo and C.~Vafa,
  arXiv:hep-th/0702077;
  R.~Tatar and B.~Wetenhall,
  JHEP {\bf 0702}, 020 (2007)
  [arXiv:hep-th/0611303];
  S.~Hirano,
  JHEP {\bf 0705}, 064 (2007)
  [arXiv:hep-th/0703272];
  A.~Giveon and D.~Kutasov,
  arXiv:hep-th/0703135;
  S.~Franco, A.~Hanany, D.~Krefl, J.~Park, A.~M.~Uranga and D.~Vegh,
  arXiv:0707.0298 [hep-th];
  E.~Dudas, J.~Mourad and F.~Nitti,
  arXiv:0706.1269 [hep-th];
  C.~Angelantonj and E.~Dudas,
  arXiv:0704.2553 [hep-th];]
  J.~Marsano, K.~Papadodimas and M.~Shigemori,
  arXiv:0705.0983 [hep-th].




\bibitem{Saltman:2004sn}
  A.~Saltman and E.~Silverstein,
  JHEP {\bf 0411}, 066 (2004)
  [arXiv:hep-th/0402135];
  M.~Gomez-Reino and C.~A.~Scrucca,
  JHEP {\bf 0605}, 015 (2006)
  [arXiv:hep-th/0602246];
  O.~Lebedev, H.~P.~Nilles and M.~Ratz,
  Phys.\ Lett.\  B {\bf 636}, 126 (2006)
  [arXiv:hep-th/0603047];
  Z.~Lalak, O.~J.~Eyton-Williams and R.~Matyszkiewicz,
  JHEP {\bf 0705}, 085 (2007)
  [arXiv:hep-th/0702026].


\bibitem{Dudas:2006gr}
  E.~Dudas, C.~Papineau and S.~Pokorski,
  JHEP {\bf 0702}, 028 (2007)
  [arXiv:hep-th/0610297];

\bibitem{Abe:2006xp}
  H.~Abe, T.~Higaki, T.~Kobayashi and Y.~Omura,
  Phys.\ Rev.\  D {\bf 75}, 025019 (2007)
  [arXiv:hep-th/0611024].

\bibitem{Kallosh:2006dv}
  R.~Kallosh and A.~Linde,
  JHEP {\bf 0702}, 002 (2007)
  [arXiv:hep-th/0611183].

\bibitem{Lebedev:2006qc}
  O.~Lebedev, V.~Lowen, Y.~Mambrini, H.~P.~Nilles and M.~Ratz,
  JHEP {\bf 0702}, 063 (2007)
  [arXiv:hep-ph/0612035].

\bibitem{Polonyi:1977pj}
  J.~Polonyi,
{\it  Hungary Central Inst Res - KFKI-77-93 (77,REC.JUL 78) 5p}.


\bibitem{O'Raifeartaigh:1975pr}
  L.~O'Raifeartaigh,
  Nucl.\ Phys.\  B {\bf 96}, 331 (1975).



\bibitem{Burgess:2003ic}
  C.~P.~Burgess, R.~Kallosh and F.~Quevedo,
  JHEP {\bf 0310}, 056 (2003)
  [arXiv:hep-th/0309187];
  A.~Achucarro, B.~de Carlos, J.~A.~Casas and L.~Doplicher,
  JHEP {\bf 0606}, 014 (2006)
  [arXiv:hep-th/0601190];
  S.~L.~Parameswaran and A.~Westphal,
  JHEP {\bf 0610}, 079 (2006)
  [arXiv:hep-th/0602253];
  arXiv:hep-th/0701215.

\bibitem{Choi:2006bh}
  K.~Choi and K.~S.~Jeong,
  JHEP {\bf 0608}, 007 (2006)
  [arXiv:hep-th/0605108].




\bibitem{Westphal:2006tn}
  A.~Westphal,
  JHEP {\bf 0703}, 102 (2007)
  [arXiv:hep-th/0611332].



\bibitem{Florea:2006si}
  B.~Florea, S.~Kachru, J.~McGreevy and N.~Saulina,
  JHEP {\bf 0705}, 024 (2007)
  [arXiv:hep-th/0610003];
  R.~Blumenhagen, M.~Cvetic and T.~Weigand,
  Nucl.\ Phys.\  B {\bf 771} (2007) 113
  [arXiv:hep-th/0609191];
  L.~E.~Ibanez and A.~M.~Uranga,
  JHEP {\bf 0703}, 052 (2007)
  [arXiv:hep-th/0609213];
  M.~Buican, D.~Malyshev, D.~R.~Morrison, H.~Verlinde and M.~Wijnholt,
  JHEP {\bf 0701}, 107 (2007)
  [arXiv:hep-th/0610007].

\bibitem{Affleck:1983mk}
  I.~Affleck, M.~Dine and N.~Seiberg,
  Nucl.\ Phys.\  B {\bf 241}, 493 (1984).


\bibitem{Acharya:2007rc}
  B.~S.~Acharya, K.~Bobkov, G.~L.~Kane, P.~Kumar and J.~Shao,
  arXiv:hep-th/0701034.

\bibitem{Abe:2006xi}
  H.~Abe, T.~Higaki and T.~Kobayashi,
  Phys.\ Rev.\  D {\bf 74}, 045012 (2006)
  [arXiv:hep-th/0606095].

\bibitem{Abe:2005rx}
  H.~Abe, T.~Higaki and T.~Kobayashi,
  Phys.\ Rev.\  D {\bf 73}, 046005 (2006)
  [arXiv:hep-th/0511160].

\bibitem{Serone:2007sv}
  M.~Serone and A.~Westphal,
  arXiv:0707.0497 [hep-th].

\bibitem{Abe:2005pi}
  H.~Abe, T.~Higaki and T.~Kobayashi,
  Nucl.\ Phys.\  B {\bf 742}, 187 (2006)
  [arXiv:hep-th/0512232].


\bibitem{Dine:2006ii}
  M.~Dine, R.~Kitano, A.~Morisse and Y.~Shirman,
  Phys.\ Rev.\  D {\bf 73}, 123518 (2006)
  [arXiv:hep-ph/0604140].


\bibitem{Kitano:2006wz}
  R.~Kitano,
  Phys.\ Lett.\  B {\bf 641}, 203 (2006)
  [arXiv:hep-ph/0607090].

\bibitem{Westphal:2007xd}
  A.~Westphal,
  arXiv:0705.1557 [hep-th].




\bibitem{Kallosh:2004yh}
  R.~Kallosh and A.~Linde,
  JHEP {\bf 0412}, 004 (2004)
  [arXiv:hep-th/0411011].

\bibitem{de Alwis:2007qx}
  S.~P.~de Alwis,
  arXiv:hep-th/0703247.

\bibitem{Wells:2004di}
  J.~D.~Wells,
  Phys.\ Rev.\  D {\bf 71}, 015013 (2005)
  [arXiv:hep-ph/0411041].
  M.~Ibe, T.~Moroi and T.~T.~Yanagida,
  Phys.\ Lett.\  B {\bf 644}, 355 (2007)
  [arXiv:hep-ph/0610277].

\bibitem{Bagger:1999rd}
  J.~A.~Bagger, T.~Moroi and E.~Poppitz,
  JHEP {\bf 0004}, 009 (2000)
  [arXiv:hep-th/9911029];
  P.~Binetruy, M.~K.~Gaillard and B.~D.~Nelson,
  Nucl.\ Phys.\  B {\bf 604}, 32 (2001)
  [arXiv:hep-ph/0011081].

\bibitem{Choi:2007ka}
  K.~Choi and H.~P.~Nilles,
  JHEP {\bf 0704}, 006 (2007)
  [arXiv:hep-ph/0702146].



\bibitem{Baer:2006id}
  H.~Baer, E.~K.~Park, X.~Tata and T.~T.~Wang,
  JHEP {\bf 0608}, 041 (2006)
  [arXiv:hep-ph/0604253]; 
%
  K.~Choi, K.~Y.~Lee, Y.~Shimizu, Y.~G.~Kim and K.~i.~Okumura,
  JCAP {\bf 0612}, 017 (2006)
  [arXiv:hep-ph/0609132]; 
%
  H.~Baer, E.~K.~Park, X.~Tata and T.~T.~Wang,
  arXiv:hep-ph/0703024; 
%
  H.~Abe, Y.~G.~Kim, T.~Kobayashi and Y.~Shimizu,
  arXiv:0706.4349 [hep-ph].




\end{thebibliography}
\end{document}